\documentstyle[prd,aps,epsfig]{revtex}

\begin{document}
%
%
\def\ov{\over}
\def\l{\left}
\def\r{\right}
\def\be{\begin{equation}}
\def\ee{\end{equation}}
\def \der#1#2{{\partial{#1}\over\partial{#2}}}
\def \dder#1#2{{\partial^2{#1}\over\partial{#2}^2}}
\def\R{{\rm I\! R}}
\def\spose#1{\hbox to 0pt{#1\hss}}
\def\lta{\mathrel{\spose{\lower 3pt\hbox{$\mathchar"218$}}
     \raise 2.0pt\hbox{$\mathchar"13C$}}}
\def\gta{\mathrel{\spose{\lower 3pt\hbox{$\mathchar"218$}}
     \raise 2.0pt\hbox{$\mathchar"13E$}}}
\draft
\title{Numerical approach for high precision 3-D relativistic star models} 
\author{Silvano Bonazzola, Eric Gourgoulhon and Jean-Alain Marck}
\address{D\'epartement d'Astrophysique Relativiste et de Cosmologie \\
  UPR 176 du C.N.R.S., Observatoire de Paris, \\
  F-92195 Meudon Cedex, France \\
  {\sl e-mail: bona@mesiob.obspm.fr, Eric.Gourgoulhon@obspm.fr,
	Jean-Alain.Marck@obspm.fr}
}
\date{8 June 1998}
\maketitle

\begin{abstract}

A multi-domain spectral method for computing very high precision 3-D
stellar models is presented. The boundary of each domain is chosen in
order to coincide with a physical discontinuity (e.g. the star's
surface).  In addition, a regularization procedure is introduced to
deal with the infinite derivatives on the boundary that may appear in
the density field when stiff equations of state are used.  Consequently
all the physical fields are smooth functions on each domain and the
spectral method is absolutely free of any Gibbs phenomenon, which
yields to a very high precision.  The power of this method is
demonstrated by direct comparison with analytical solutions such as
MacLaurin spheroids and Roche ellipsoids.  The relative numerical error
reveals to be of the order of $10^{-10}$. This approach has been
developed for the study of relativistic inspiralling binaries. It may
be applied to a wider class of astrophysical problems such as the study
of relativistic rotating stars too.

\end{abstract}


\section{Introduction}

One of the most promising source of gravitational waves is the
coalescence of inspiralling compact binaries. The recent development of
interferometric gravitational waves detectors ({\sl e.g. GEO600, LIGO,
TAMA} and {\sl VIRGO}) 
gives an important motivation for studying this problem.
Such a study requires a relativistic formalism to derive the equations
of motion and then an accurate and tricky method to solve the resulting
system of partial differential equations. We have recently
\cite{BonGM97a} proposed a relativistic formalism able to tackle the
problem of co-rotating {\sl as well as} counter-rotating binaries
system, these latter being more relevant from the astrophysical point of view.
We present now a very accurate approach based on multi-domain
spectral method that circumvents the Gibbs phenomenon to numerically
solve this problem and which can be applied to a wide class of other
astrophysical situations.

Various astrophysical applications of spectral methods have been
developed in our group (for a review, see \cite{BonGM97b}), including
3-D gravitational collapse of stellar core \cite{BonM93}, neutron star
collapse into a black hole \cite{Gou91,GouH93,GouHG95,Nov98}, 
tidal disruption of a star near a
massive black hole \cite{Mar96}, rapidly rotating neutron stars
\cite{BonGSM93,SalBGH94a,SalBGH94b,HaeBS95},
magnetized neutron stars \cite{BocBGN95,HaeB96} and their resulting
gravitational radiation \cite{BonG96}, spontaneous symmetry breaking of
rapidly rotating neutron stars \cite{BonFG96,BonFG98},
proto-neutron stars evolution \cite{Gou97,GouHZ97,GouHZ98}.

In computational fluid dynamics, spectral methods are known for their
very high accuracy \cite{GotO77,CanHQZ88};  indeed for a ${\cal
C}^\infty$ function, the numerical error decreases as $\exp(-N)$ ({\sl
evanescent error}), where $N$ is the number of coefficients involved in
the spectral expansion, or equivalently the number of grid points in
the physical domain. This is much faster than the error decay of
finite-difference methods, which behaves as $1/N^q$, with $q$ generally
not larger than $3$.  For this reason, spectral methods are particularly
interesting for treating 3-D problems --- such as binary configurations
--- a situation in which the number of grid points is still severely
limited by the capability of present and next generation computers.

Spectral methods lose much of their accuracy when non-smooth functions
are treated because of the so-called {\em Gibbs phenomenon}. This
phenomenon is well known from the most familiar spectral method, namely
the theory of Fourier series: the Fourier coefficients $(c_n)$ of a
function $f$ which is of class ${\cal C}^p$ but not ${\cal C}^{p+1}$
decrease as $1/n^p$ only. In particular, if the function has some
discontinuity, its approximation by a Fourier series does not converge
towards $f$ at the discontinuity point: there remains a gap which is of
the order 10\%.

The multi-domain spectral method described in this paper circumvents
the Gibbs phenomenon. The basic idea is to divide the space into
domains chosen so that the physical discontinuities are located onto
the boundaries between the domains (Sect.~\ref{s:mapping}).
The simplest example is the case of
a perfect fluid star, where two domains may be distinguished:  the
interior and the exterior of the star. The boundary is then simply the
surface of the star. The second ingredient of the technique is a
mapping between the domains defined in this way and some simple
mathematical domains, which are cross products of intervals:
$[a_1,a_2] \times [b_1,b_2] \times [c_1,c_2]$. The spectral expansion
is then performed with respect to functions of the coordinates spanning
these intervals (Sect.~\ref{s:multi}). The method of resolution of a basic
equation, namely the Poisson equation, is exposed in Sect.~\ref{s:resol_Poisson}. 
For stiff equations of state, the above procedure is not sufficient to
ensure the smoothness of all the functions.  Indeed, for a polytrope
with an adiabatic index greater than 2, the density field has an
infinite derivative on the surface of the star. We present in
Sect.~\ref{s:regul} a method for regularizing the density and recover
the spectral precision. The power of the multi-domain spectral 
method is illustrated in Sect.~\ref{s:illustr}, where comparisons are
performed between numerical solutions obtained by an implementation of the
method and analytical solutions (ellipsoidal configurations of incompressible
fluids). Finally Sect.~\ref{s:conclu} concludes the article by discussing
the great advantages of the multi-domain spectral method for dealing with
relativistic binary neutron stars.

\section{The physical domains and their mapping} \label{s:mapping}

\subsection{Splitting of the physical space into star-like domains}

\begin{figure}
\centerline{ \epsfig{figure=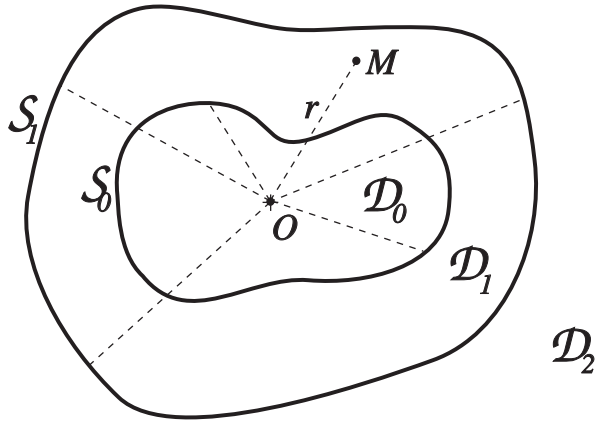,height=6cm} }
\caption[]{\label{f:domaine} 
Splitting of the physical three-dimensional space into domains ${\cal D}_0$, 
${\cal D}_1$, ...,${\cal D}_{{\cal N}-1}$ (on the figure ${\cal N}=3$),
which are star-like with respect to some origin $O$. The last domain
(here ${\cal D}_2$) extends up to infinity.}  
\end{figure}

In order to treat problems involving Poisson equations with non-compact
sources --- as they appear in relativistic gravitation --- we take for
the physical domain where the computation must be performed the whole
three-dimensional space $\R^3$. In doing so, we know the physical boundary
conditions we have to impose in order to solve the Poisson equations. 
These boundary conditions can be easily set at infinity.  
We divide $\R^3$
into ${\cal N}$ domains $({\cal D}_l)_{0\leq l\leq {\cal N}-1}$ (${\cal
N}\geq 2$). In the present work, these domains are taken to be {\em
star-like} (in the mathematical sense)
with respect to a some origin $O$, which means that for every
point $M$ in the domain ${\cal D}_l$, the segment $OM$ is entirely
included in 
$ \bigcup_{i \leq l} {\cal D}_l $
(see Fig.~\ref{f:domaine}). The multi-domain spectral method we are
going to describe can be extended to more general domains, at the price
of a greater technical (but non conceptual) difficulty. However, for
stellar configurations, the star-like hypothesis is sufficient for most
applications.  Let us denote by ${\cal S}_l$ the boundary surface
between the domains ${\cal D}_l$ and ${\cal D}_{l+1}$. ${\cal D}_0$ is
simply connected and its boundary is ${\cal S}_0$; we call it the {\em
nucleus}.  For $1 \leq l \leq {\cal N}-2$, ${\cal D}_l$'s inner
boundary is ${\cal S}_{l-1}$ and outer boundary ${\cal S}_l$. The last
domain, ${\cal D}_{{\cal N}-1}$, has ${\cal S}_{{\cal N}-2}$ as inner
boundary and extends to infinity (cf. Fig.~\ref{f:domaine}).

Let us choose some Cartesian frame of $\R^3$ centered at $O$ and let us
call $(r,\theta,\varphi)$ the associated spherical coordinates:  $r\in
[0,+\infty[$, $\theta\in [0,\pi]$ and $\varphi\in [0, 2\pi[$.  The
mapping of each domain onto the cross product of intervals $[a_1,a_2]
\times [b_1,b_2] \times [c_1,c_2]$ will be defined with respect to the
spherical coordinates $(r,\theta,\varphi)$.  Since each domain ${\cal
D}_l$ is star-like with respect to $O$, the equation of the boundaries
${\cal S}_l$ can be written under the form 
\be \label{e:S_l:def}
	r = S_l(\theta,\varphi) \ , 
\ee 
where $S_l$ is a smooth function on $[0,\pi]\times [0,2\pi[$.

\subsection{Mapping of the nucleus} \label{s:map_nucl}

The basic idea is to introduce a mapping
\be \label{e:map:gen}
	\begin{array}{rcl}
  [0,1]\times[0,\pi]\times[0,2\pi[ & \longrightarrow & {\cal D}_0 \\
    (\xi,\theta',\varphi') & \longmapsto & (r,\theta,\varphi)
	\end{array}
\ee
so that the origin $O$ corresponds to $\xi=0$ and the boundary
${\cal S}_0$ to $\xi = {\rm const} = 1$. Using the fact that ${\cal D}_0$
is star-like, a simple form of the mapping (\ref{e:map:gen}) 
can be chosen as 
\begin{eqnarray}
    r & = & R_0(\xi,\theta',\varphi') \label{e:r=R0} \\
    \theta & = & \theta' \label{e:theta} \\
    \varphi & = & \varphi' \label{e:phi} \ ,
\end{eqnarray}
where $R_0$ is a smooth function subject to regularity properties which
are discussed below. Thanks to Eqs.~(\ref{e:theta})-(\ref{e:phi}), we
will make no distinction between $\theta$ and $\theta'$, as well as 
between $\varphi$ and $\varphi'$, i.e. we will abandon the primes on the
angles.
The fact that ${\cal D}_0$'s boundary coincides with $\xi=1$ translates
as
\be \label{e:R0=surf}
	R_0(1,\theta,\varphi) = S_0(\theta,\varphi) \ .
\ee
Beside Eq.~(\ref{e:R0=surf}), the function $R_0$ must satisfy some 
{\em regularity properties} induced by the singular behavior of spherical
coordinates at $r=0$, $\theta=0$ and $\theta=\pi$.
We define a function $f: \R^3 \rightarrow \R$ to be {\em regular}
if it can be expressed as a polynomial of the Cartesian coordinates
\begin{eqnarray}
 x & = & r \sin\theta \cos\varphi \label{e:cart,x} \\
 y & = & r \sin\theta \sin\varphi \\
 z & = & r \cos\theta \label{e:cart,z} \ .
\end{eqnarray}
We will assume that all the physical fields are
regular functions on each domain ${\cal D}_l$ (the domains ${\cal D}_l$
are in fact chosen in this manner) with respect to the previous 
definition. 
It is easy to see that any regular function is expandable as
\be \label{e:devel,regul}
  f(r,\theta,\varphi) = \sum_{m=-M}^{M} \ 
	\sum_{\ell = |m|}^{L}
  r^\ell T(r^2) 
\sin^{|m|}\theta \ \  P_{\ell-|m|}(\cos\theta) \  e^{im\varphi} \ ,
\ee
where $L$ and $M$ are positive integers, $L \geq M$,  
$P_{\ell-|m|}$ is some polynomial of degree $\ell-|m|$ and $T(r^2)$ is
some even polynomial.

A simple form of the mapping 
(\ref{e:r=R0})-(\ref{e:phi}) has been already introduced in the literature
\cite{EriM85,EriM91,UryE96}, namely 
$R_0(\xi,\theta,\varphi) = S_0(\theta,\varphi)\, \xi$, where 
$S_0(\theta,\varphi)$ is the equation of the star's surface
[Eq.~(\ref{e:S_l:def})]. However for such a mapping the regularity
condition (\ref{e:devel,regul}) would be quite complicated when expressed
in terms of $(\xi,\theta,\varphi)$. We choose instead the mapping defined by
\be \label{e:r0=afbg}
  R_0(\xi,\theta,\varphi) = \alpha_0 \l[ \xi + A_0(\xi) F_0(\theta,\varphi)
	+ B_0(\xi) G_0(\theta,\varphi) \r] \ ,
\ee
where $A_0$ and $B_0$ are the following even and odd 
polynomials
\begin{eqnarray}
  A_0(\xi) & = & 3 \xi^4 - 2 \xi^6 \label{e:A0:def} \\
  B_0(\xi) & = & (5\xi^3 - 3\xi^5) / 2 \label{e:B0:def} \ ,
\end{eqnarray}
and the constant $\alpha_0$ as well as the two functions
$F_0(\theta,\varphi)$ and $G_0(\theta,\varphi)$ are such that (i)
the Fourier expansion of $F_0(\theta,\varphi)$ (resp. 
$G_0(\theta,\varphi)$) with respect to $\varphi$ contains only
even (resp. odd) harmonics and (ii) the 
equation of the surface ${\cal S}_0$ can be written
\be
  \alpha_0 \l[ 1 + F_0(\theta,\varphi) + G_0(\theta,\varphi) \r] =
	S_0(\theta,\varphi) \label{e:f0g0=s0} \ .
\ee

The polynomials $A_0$ and $B_0$ defined by 
(\ref{e:A0:def})-(\ref{e:B0:def}) are such that
\begin{eqnarray}
   & & A_0(0) = B_0(0) = 0 \\
   & & A_0'(0) = B_0'(0) = 0 \\
   & & A_0(1) = B_0(1) = 1 \label{e:a0b0=1} \\
   & & A_0'(1) = B_0'(1) = 0 \ .   
\end{eqnarray}
The properties (\ref{e:f0g0=s0}) and (\ref{e:a0b0=1}) ensure that
Eq.~(\ref{e:R0=surf}) is satisfied, i.e. that the mapping 
(\ref{e:r0=afbg}) is from $[0,1]\times[0,\pi]\times[0,2\pi[$ to
${\cal D}_0$. The polynomials $A_0$ and $B_0$ are chosen
in order to satisfy at the minimum level the regularity conditions
mentioned above. In particular, near the origin $r$ behaves as
$r \sim \alpha_0 \xi$ and is independent of $(\theta,\varphi)$, which 
would not have been the case of the mapping $r=S_0(\theta,\varphi)\, \xi$
introduced in Refs.~\cite{EriM85,EriM91,UryE96}.

The Jacobian of
the transformation $(r,\theta,\varphi) \mapsto (\xi,\theta,\varphi)$
is 
\begin{eqnarray}
   J & := & {\partial(r,\theta,\varphi) \ov \partial(\xi,\theta,\varphi) }
   	=  \der{R_0}{\xi} \nonumber \\
   &=& \alpha_0 \l[ 1 + A_0'(\xi) F_0(\theta,\varphi)
		+ B_0'(\xi) G_0(\theta,\varphi) \r]
\end{eqnarray}
Since $A_0'(\xi) \geq 0$ and $B_0'(\xi)\geq 0$ for any $\xi \in [0,1]$,
the mapping could become singular if $F_0(\theta,\varphi)$ or
$G_0(\theta,\varphi)$ is negative and has an important amplitude.
We cannot control the sign of the function $F_0(\theta,\varphi)$ because
it contains only odd harmonics of $\varphi$. However, by a suitable
choice of $\alpha_0$, one can impose
\be \label{e:g0>0}
    G_0(\theta,\varphi) \geq 0 \ ,
\ee
as we shall see below.
We have found that this condition was sufficient to ensure that $J\not =0$,
i.e. that the mapping is regular, for all the astrophysically relevant 
situations we have encountered.

The equation for the surface of the domain ${\cal D}_0$ being given, under 
the form (\ref{e:S_l:def}), $r=S_0(\theta,\varphi)$, the procedure
which leads to $\alpha_0$, $F_0(\theta,\varphi)$ and $G_0(\theta,\varphi)$,
i.e. to the full determination of the function $R_0(\xi,\theta,\varphi)$,
is the following one. 

First let us choose a point  
$(\theta_*,\varphi_*)$ on the surface ${\cal S}_0$. 
Equation~(\ref{e:f0g0=s0}) implies the following relation
\be \label{e:s0(*):1}
  \alpha_0 \l[ 1 + F_0(\theta_*,\varphi_*) + G_0(\theta_*,\varphi_*) \r] =
	S_0(\theta_*,\varphi_*) \ .
\ee
Let us introduce the following auxiliaries quantities
\begin{eqnarray}
  & & \mu := \alpha_0 \left[
		F_0(\theta_*,\varphi_*) 
		+ G_0(\theta_*,\varphi_*) 
		\right] \\
  & & \tilde F_0(\theta,\varphi) := \alpha_0 F_0(\theta,\varphi)
			\label{e:tf0:def} \\
  & & \tilde G_0(\theta,\varphi) := \alpha_0 G_0(\theta,\varphi)
	- \mu 		\label{e:tg0:def} \ .
\end{eqnarray}
Equation (\ref{e:s0(*):1}) translates then in
\be \label{e:a+l+m=s}
  \alpha_0 + \mu = S_0(\theta_*,\varphi_*) \ ,
\ee
whereas Eq.~(\ref{e:f0g0=s0}) becomes
\be
  \tilde F_0(\theta,\varphi) + \tilde G_0(\theta,\varphi) =
	S_0(\theta,\varphi) - S_0(\theta_*,\varphi_*) \ . 
\ee
Having  expanded $f(\theta,\varphi) := S_0(\theta,\varphi) - 
S_0(\theta_*,\varphi_*)$ into Fourier series with respect to $\varphi$,
one deduces the functions $\tilde F_0(\theta,\varphi)$ (resp.
$\tilde G_0(\theta,\varphi)$) by taking only the odd (resp. even)
harmonics of this Fourier expansion. $\mu$ is then
computed as
\be
	\mu = - \min \l\{ \tilde G_0(\theta,\varphi), \ 
	(\theta,\varphi) \in [0,\pi]\times [0,2\pi[ \r\} \ . 
\ee
In doing so, condition (\ref{e:g0>0}) will 
automatically be fulfilled.
The value of the coefficient $\alpha_0$ is deduced from the above 
value of $\mu$ via Eq.~(\ref{e:a+l+m=s}). 
Finally the functions $F_0(\theta,\varphi)$ and $G_0(\theta,\varphi)$
are computed from Eqs.~(\ref{e:tf0:def}) and (\ref{e:tg0:def}).

\subsection{Mapping of the intermediate domains} \label{s:map_interm}

For $1 \leq l \leq  {\cal N}-2$, we introduce the mapping 
\be \label{e:map:interm}
	\begin{array}{rcl}
  [-1,1]\times[0,\pi]\times[0,2\pi[ & \longrightarrow & {\cal D}_l \\
    (\xi,\theta',\varphi') & \longmapsto & (r,\theta,\varphi)
	\end{array}
\ee
under the form
\begin{eqnarray}
    r & = & R_l(\xi,\theta',\varphi') \\
    \theta & = & \theta' \\
    \varphi & = & \varphi'  \ ,
\end{eqnarray}
where $R_l$ is a smooth function which satisfies
\begin{eqnarray}
 & & R_l(-1,\theta,\varphi) = S_{l-1}(\theta,\varphi) \label{e:rlm1=surf} \\
 & & R_l(+1,\theta,\varphi) = S_l(\theta,\varphi) \ , \label{e:rlp1=surf}
\end{eqnarray}
which means that the inner (resp. outer) boundary of ${\cal D}_l$ is defined 
by $\xi=-1$ (resp. $\xi=+1$).

We choose $R_l(\xi,\theta,\varphi)$ as
\begin{eqnarray}
  R_l(\xi,\theta,\varphi) & = & \alpha_l
	 \l[ \xi + A_l(\xi) F_l(\theta,\varphi)
	+ B_l(\xi) G_l(\theta,\varphi) \r] \nonumber \\
	& & + \beta_l \ , \label{e:rl=afbg}
\end{eqnarray}
where $A_l$ and $B_l$ are the following polynomials
\begin{eqnarray}
  A_l(\xi) & = & ( \xi^3 - 3 \xi + 2 ) / 4 \label{e:Al:def} \\
  B_l(\xi) & = & (- \xi^3 + 3\xi + 2 ) / 4 \label{e:Bl:def} \ ,
\end{eqnarray}
and the constants $\alpha_l$ and $\beta_l$ and the two functions
$F_l(\theta,\varphi)$ and $G_l(\theta,\varphi)$ are defined from the
equations of the surfaces ${\cal S}_{l-1}$ and ${\cal S}_l$ by
\begin{eqnarray}
  & & \alpha_l \l[ - 1 + F_l(\theta,\varphi) \r] 
	+ \beta_l = S_{l-1}(\theta,\varphi) \label{e:flglm1=slm1} \\
  & & \alpha_l \l[ + 1  + G_l(\theta,\varphi) \r] 
	+ \beta_l = S_l(\theta,\varphi) \label{e:flgl=sl} \\
  & & F_l(\theta,\varphi) \leq 0 \label{e:fl<0} \\
  & & G_l(\theta,\varphi) \geq 0 \label{e:gl>0}
\end{eqnarray}
Note that the polynomials $A_l$ and $B_l$ defined by 
(\ref{e:Al:def})-(\ref{e:Bl:def}) are such that
\begin{eqnarray}
   & & A_l(-1) = 1 \quad \mbox{and} \quad B_l(-1) = 0 \label{e:albl(0)} \\
   & & A_l(+1) = 0 \quad \mbox{and} \quad B_l(+1) = 1 \label{e:albl(1)} \\
   & & A_l'(-1) = B_l'(-1) = A_l'(+1) = B_l'(+1) = 0 \ .   
\end{eqnarray}
The properties (\ref{e:flglm1=slm1}) and (\ref{e:albl(0)}) [resp.
(\ref{e:flgl=sl}) and (\ref{e:albl(1)})] ensure that Eq.~(\ref{e:rlm1=surf})
[resp. Eq.~(\ref{e:rlp1=surf})] is satisfied, i.e. that the mapping
(\ref{e:rl=afbg}) is from $[-1,1]\times[0,\pi]\times[0,2\pi[$
to ${\cal D}_l$. The conditions (\ref{e:fl<0}) and (\ref{e:gl>0}) ensure
that this mapping is not singular, by the same argument as that 
presented for $R_0$ in Sect.~\ref{s:map_nucl}, the sign of 
$F_l(\theta,\varphi)$ being opposite to that of $G_l(\theta,\varphi)$
because $A_l$ is a decreasing function of $\xi$, whereas $B_l$ is
an increasing function of $\xi$.

The equation for the inner and outer boundaries of the domain 
${\cal D}_l$ being given, under 
the form (\ref{e:S_l:def}): $r=S_{l-1}(\theta,\varphi)$ (inner boundary)
and $r=S_l(\theta,\varphi)$ (outer boundary), the procedure
which leads to $\alpha_l$, $\beta_l$, $F_l(\theta,\varphi)$ and 
$G_l(\theta,\varphi)$,
i.e. to the full determination of the function $R_l(\xi,\theta,\varphi)$,
is the following one. 

First let us choose a point  
$(\theta_*,\varphi_*)$ on the surface ${\cal S}_{l-1}$ along with 
the corresponding point $(\theta_*,\varphi_*)$ on the surface ${\cal S}_l$.
Equations~(\ref{e:flglm1=slm1}) and (\ref{e:flgl=sl}) imply the following 
relations
\begin{eqnarray} 
  &  & \alpha_l \l[ - 1 + F_l(\theta_*,\varphi_*)  \r] + \beta_l =
	S_{l-1}(\theta_*,\varphi_*) \label{e:slm1(*):1} \\
  &  & \alpha_l \l[ 1 + G_l(\theta_*,\varphi_*)  \r] + \beta_l =
	S_l(\theta_*,\varphi_*) \label{e:sl(*):1} \ .
\end{eqnarray}
Let us introduce the following auxiliaries quantities
\begin{eqnarray}
  & & \lambda := \alpha_l F_l(\theta_*,\varphi_*) \\
  & & \mu := \alpha_l G_l(\theta_*,\varphi_*) \\
  & & \tilde F_l(\theta,\varphi) := \alpha_l F_l(\theta,\varphi)
	- \lambda \label{e:tfl:def} \\
  & & \tilde G_l(\theta,\varphi) := \alpha_l G_l(\theta,\varphi)
	- \mu 		\label{e:tgl:def} \ .
\end{eqnarray}
Equations (\ref{e:slm1(*):1}) and (\ref{e:sl(*):1}) then translate in
\begin{eqnarray}
  & & - \alpha_l + \lambda + \beta_l = S_{l-1}(\theta_*,\varphi_*) 
					\label{e:slm1(*):2} \\
  & &  \alpha_l + \mu + \beta_l = S_l(\theta_*,\varphi_*) 
					\label{e:sl(*):2} \ ,
\end{eqnarray}
whereas Eqs.~(\ref{e:flglm1=slm1}) and (\ref{e:flgl=sl}) become
\begin{eqnarray}
  &  & \tilde F_l(\theta,\varphi)  =
	S_{l-1}(\theta,\varphi) - S_{l-1}(\theta_*,\varphi_*) \\
  &  & \tilde G_l(\theta,\varphi)  =
	S_l(\theta,\varphi) - S_l(\theta_*,\varphi_*) \ .
\end{eqnarray}
From the values of $\tilde F_l(\theta,\varphi)$ and 
$\tilde G_l(\theta,\varphi)$ obtained above, 
$\lambda$ and $\mu$ are computed as
\begin{eqnarray}
   & & \lambda = - \max \l\{ \tilde F_l(\theta,\varphi), \ 
	(\theta,\varphi) \in [0,\pi]\times [0,2\pi[ \r\} \\ 
   & & \mu = - \min \l\{ \tilde G_l(\theta,\varphi), \ 
	(\theta,\varphi) \in [0,\pi]\times [0,2\pi[ \r\} \ . 
\end{eqnarray}
In doing so, the conditions (\ref{e:fl<0}) and (\ref{e:gl>0}) will 
automatically be fulfilled.
The value of the constants $\alpha_l$ and $\beta_l$ are deduced from the 
above values of $\lambda$ and $\mu$ via Eqs.~(\ref{e:slm1(*):2}) and
(\ref{e:sl(*):2}). 
Finally the functions $F_l(\theta,\varphi)$ and $G_l(\theta,\varphi)$
are computed from Eqs.~(\ref{e:tfl:def}) and (\ref{e:tgl:def}).

\subsection{Compactification of the infinite domain} \label{s:map_externe}

In the case where the external domain 
${\cal D}_{\rm ext} := {\cal D}_{{\cal N}-1}$ extends to infinity 
we introduce the mapping 
\be \label{e:map:ext}
	\begin{array}{rcl}
  [-1,1]\times[0,\pi]\times[0,2\pi[ & \longrightarrow & {\cal D}_{\rm ext} \\
    (\xi,\theta',\varphi') & \longmapsto & (r,\theta,\varphi)
	\end{array}
\ee
under the form
\begin{eqnarray}
    u & := & 1/r = U(\xi,\theta',\varphi') \\
    \theta & = & \theta' \\
    \varphi & = & \varphi'  \ ,
\end{eqnarray}
where $U$ is a smooth function which satisfies
\begin{eqnarray}
 & & U(-1,\theta,\varphi) = S_{\rm ext}(\theta,\varphi)^{-1} 
					\label{e:um1=surf} \\
 & & U(+1,\theta,\varphi) = 0 \ , \label{e:up1=surf}
\end{eqnarray}
where $S_{\rm ext}(\theta,\varphi) := S_{{\cal N}-2}(\theta,\varphi)$.
The above two equations show
 that the inner boundary of ${\cal D}_{\rm ext}$ is defined 
by $\xi=-1$, whereas $\xi=+1$ corresponds to the infinity.
We have already introduced such a compactification of the infinite domain
in Ref.~\cite{BonGSM93}, in the case of a spherical inner boundary. 

We choose the function $U(\xi,\theta,\varphi)$ as
\be
  U(\xi,\theta,\varphi) =  \alpha_{\rm ext}
	 \l[ \xi + A_{\rm ext}(\xi) F_{\rm ext}(\theta,\varphi) - 1 \r]
						 \ , \label{e:u=af}
\ee
where $A_{\rm ext}$ is the same polynomial of $\xi$ as that 
defined in Eq.~(\ref{e:Al:def}), and the constant 
$\alpha_{\rm ext}$ and the function
$F_{\rm ext}(\theta,\varphi)$ are defined from the
equations of the surface ${\cal S}_{{\cal N}-2}$ by
\begin{eqnarray}
  & & \alpha_{\rm ext} \l[ - 2 + F_{\rm ext}(\theta,\varphi) \r] 
	 = S_{\rm ext}(\theta,\varphi)^{-1} \label{e:fext=s} \\
  & & F_{\rm ext}(\theta,\varphi) \leq 0 \label{e:fext<0} \ .
\end{eqnarray}
The condition (\ref{e:fext<0}) ensures that 
$\partial U/\partial\xi \not = 0$
i.e. that the mapping (\ref{e:u=af})
is not singular.

The equation for the inner boundary of the domain ${\cal D}_{\rm ext}$ 
being given, under 
the form (\ref{e:S_l:def}): $r=S_{\rm ext}(\theta,\varphi)$, the procedure
which leads to $\alpha_{\rm ext}$ and $F_{\rm ext}(\theta,\varphi)$,
i.e. to the full determination of the function $U(\xi,\theta,\varphi)$,
is the following one. First let us choose a point  
$(\theta_*,\varphi_*)$ on the surface ${\cal S}_{\rm ext}$. 
Equation~(\ref{e:fext=s}) implies the following relation
\be 
  \alpha_{\rm ext} \l[ -2 + F_{\rm ext}(\theta_*,\varphi_*) \r] =
	S_{\rm ext}(\theta_*,\varphi_*)^{-1} \ .
\ee
By introducing the auxiliaries quantities
\begin{eqnarray}
  & & \lambda := \alpha_{\rm ext} F_{\rm ext}(\theta_*,\varphi_*) \\
  & & \tilde F_{\rm ext}(\theta,\varphi) := \alpha_{\rm ext}
	 F_{\rm ext}(\theta,\varphi)
	- \lambda \label{e:tfext:def} \ , 
\end{eqnarray}
this equation translates in
\be \label{e:a+l=s:ext}
  - 2 \alpha_{\rm ext} + \lambda  = S_{\rm ext}(\theta_*,\varphi_*)^{-1} \ ,
\ee
whereas Eq.~(\ref{e:fext=s}) becomes
\be
  \tilde F_{\rm ext}(\theta,\varphi)  =
	S_{\rm ext}(\theta,\varphi)^{-1} - 
	S_{\rm ext}(\theta_*,\varphi_*)^{-1} \ . 
\ee
From the above value of $\tilde F_{\rm ext}(\theta,\varphi) $,
 $\lambda$ is computed according to
\be
  \lambda = - \min \l\{ \tilde F_{\rm ext}(\theta,\varphi), \ 
	(\theta,\varphi) \in [0,\pi]\times [0,2\pi[ \r\} \ . 
\ee
In doing so, the condition (\ref{e:fext<0}) will 
automatically be fulfilled (recall that $\alpha_{\rm ext} <0$).
The value of $\alpha_{\rm ext}$ is deduced from the above 
value of $\lambda$ via Eq.~(\ref{e:a+l=s:ext}). 
Finally the function $F_{\rm ext}(\theta,\varphi)$ is
computed from Eq.~(\ref{e:tfext:def}).

\section{Multi-domain spectral method} \label{s:multi}
 
\subsection{Spectral expansion of a physical field}	\label{s:expansion}

The spirit of the multi-domain spectral method is to perform 
spectral expansions on each domain ${\cal D}_l$, and
with respect to the coordinates $(\xi,\theta,\varphi)$
instead of the physical coordinates $(r,\theta,\varphi)$. 
We shall take as basis functions {\em separable} functions of 
$(\xi,\theta,\varphi)$, i.e. functions that can be put under the
form $X(\xi) \Theta(\theta) \Phi(\varphi)$.
The variable $\varphi$ being periodic, it is natural to use 
Fourier series in $\varphi$, i.e. to choose 
\be
  \Phi_k(\varphi) = e^{ik\varphi} \quad 
	{- N_\varphi/2 \leq k \leq N_\varphi/2} \ ,
\ee
where $N_\varphi$ is an even integer that we will call the {\em number of
degrees of freedom in $\varphi$}. 
The associated collocation points (``grid points'') are
\be
  \varphi_k = 2 \pi \, k / N_\varphi \, \quad 0 \leq k \leq N_\varphi - 1 \ .
\ee
As concerns $\Theta(\theta)$ one must use functions that are compatible
with the expansion (\ref{e:devel,regul}) of any regular scalar field $f$.
We shall not use $\sin^{|m|}\theta \  P_{\ell-|m|}(\cos\theta)$,
as suggested by Eq.~(\ref{e:devel,regul}), but a wider set, namely the
functions
\begin{eqnarray}
   & & 	\Theta_{kj}(\theta) = \cos(j \theta) \quad 0 \leq j \leq N_\theta - 1 
	\quad \mbox{for $m$ even} 
					\label{e:cosjt} \\
   & & 	\Theta_{kj}(\theta) = \sin(j \theta)  \quad 1 \leq j \leq N_\theta - 2 
	\quad \mbox{for $m$ odd} 
					\label{e:sinjt} \ ,
\end{eqnarray}
where $N_\theta$ is an integer that we will call the {\em number of
degrees of freedom in $\theta$} and
$m$ is the degree of the harmonic in the Fourier series with 
respect to $\varphi$: $m=k$ in the present case.
The advantages of this choice are to allow the use of 
Fast-Fourier-Transform algorithms for computing the coefficients,
as well as very simple matrices for the usual differential operators
\cite{BonGM98}. 
The associated collocation points  are
\be
  \theta_j = \pi \, j / (N_\theta-1) \, \quad 0 \leq j \leq N_\theta - 1 \ .
\ee

As concerns the variable $\xi$, we also choose a set wider than 
merely $\xi^\ell$, namely:
\begin{eqnarray}
  & & X_{kji}(\xi) = T_{2i}(\xi)  \quad 0 \leq i \leq N_r - 1 
					\quad \mbox{for $j$ even} \\	
  & & X_{kji}(\xi) = T_{2i+1}(\xi)  \quad 0 \leq i \leq N_r - 2
						\quad \mbox{for $j$ odd} \ ,	
\end{eqnarray}
where $N_r$ is an integer that we will call the {\em number of
degrees of freedom in $r$} and 
$T_n$ denotes the $n^{\rm th}$ degree Chebyshev polynomial.
The associated collocation points  are
\be \label{e:xi:coll:nucl}
  \xi_i = \sin \l( {\pi \ov 2} {i\ov N_r -1} \r)
			 \, \quad 0 \leq i \leq N_r - 1 \ .
\ee

The above choice concerns the nucleus ${\cal D}_0$ only. For the
intermediate and external domains, we choose instead
\be
   X_{kji}(\xi) = T_i(\xi) \ ,
\ee
along with the collocation points
\be
  \xi_i = - \cos (\pi \,   i /( N_r -1) ) 
			 \, \quad 0 \leq i \leq N_r - 1 \ .
\ee

Note that for the nucleus the above choice is the same as that
presented in \cite{BonM90}, once $\xi$ is replaced by $r$. We report
the interested reader to that paper for a more detailed discussion
about this choice (see also Appendices A, B and D of \cite{Gou97}).

When a symmetry is present, we use different bases, in order to take
the symmetry into account. For instance, an often existing symmetry
is the symmetry with respect to the equatorial plane, i.e. the
plane $\theta = \pi / 2$. In this case, we use, instead of 
(\ref{e:cosjt})-(\ref{e:sinjt}),
\begin{eqnarray}
   & & 	\Theta_{kj}(\theta) = \cos(2j \theta) \quad \mbox{for $m$ even} 
					\label{e:cos2jt} \\
   & & 	\Theta_{kj}(\theta) = \sin((2j+1) \theta) \quad \mbox{for $m$ odd} 
					\label{e:sin2jt} \ .
\end{eqnarray}
The associated collocation points span only $[0,\pi/2]$, instead
of $[0,\pi]$:
\be \label{e:theta:col:sym}
  \theta_j = {\pi\ov 2} \, {j \ov N_\theta-1} \, 
				\quad 0 \leq j \leq N_\theta - 1 \ .
\ee

Another usual symmetry is the above equatorial symmetry augmented by
the symmetry under the transformation $\varphi \mapsto \varphi + \pi$.
This is the case of a triaxial ellipsoid, or of an axisymmetric star
perturbed by even $m$ modes.
In this case, the $\varphi$-basis functions are 
\be
  \Phi_k(\varphi) = e^{2ik\varphi} \ .
\ee
The associated collocation points span $[0,\pi[$, instead of
$[0,2\pi[$:
\be
  \varphi_k = \pi \, k / N_\varphi \, \quad 0 \leq k \leq N_\varphi - 1 \ .
\ee
The basis in $\theta$ become
\be
   \Theta_{kj}(\theta) = \cos(2j \theta)  \ ,
\ee
instead of (\ref{e:cos2jt})-(\ref{e:sin2jt}), the collocation points in
$\theta$ remaining those given by Eq.~(\ref{e:theta:col:sym}).
In this case, the basis for $\xi$ in the nucleus contain only 
even polynomials:
\be
   X_{kji}(\xi) = T_{2i}(\xi) \ ,
\ee
the collocation points remaining those given by Eq.~(\ref{e:xi:coll:nucl}).

\subsection{Differential operators}

In this section, we present how a first order
differential operator, the gradient, and a second order one, the
Laplacian, both applied to a scalar field, are expressed in term of the 
coordinates system described above. The computation of any other kind
of operator is straightforward.

The components of the gradient of a scalar field $f$ in an orthonormal
basis associated with the spherical coordinates $(r,\theta,\varphi)$
are
\begin{eqnarray}
 \der{f}{r} & = & J_1^{-1} \der{f}{\xi} \\
 {1\ov r} \der{f}{\theta} & = & {1\ov R_l} \der{f}{\theta'} -
	{J_2 \ov J_1} \der{f}{\xi} \\
 {1\ov r\sin\theta} \der{f}{\varphi} & = & {1\ov R_l\sin\theta'} 
	\der{f}{\varphi'} - {J_3 \ov J_1} \der{f}{\xi} \ , 	
\end{eqnarray}
where the following abbreviations have been introduced:
\begin{eqnarray}
  J_1 & := & \der{R_l}{\xi} \\
  J_2 & := & {1\ov R_l}\der{R_l}{\theta'} \\
  J_3 & := & {1\ov R_l\sin\theta'}\der{R_l}{\varphi'} \ .
\end{eqnarray}
Note that we have re-introduced the primes on $\theta$ and $\varphi$
[cf. Eqs. (\ref{e:theta})-(\ref{e:phi})]
to avoid any confusion between the partial derivatives.
The partial derivatives that appear in the quantities $J_i$ are
computed by (i) a (banded) matrix multiplication on the
coefficients of the spectral expansion of the functions 
$F_l(\theta,\varphi)$ and $G_l(\theta,\varphi)$ and (ii) analytically
for the polynomials $A_l(\xi)$ and $B_l(\xi)$. In the nucleus,
$J_2$ is re-expressed as
\be
  J_2  =  {(3\xi^3 -2\xi^5) \partial F_0/\partial\theta' + {1\ov 2} 
	(5\xi^2 - 3 \xi^4)  \partial G_0/\partial\theta'  \ov
	1 + (3\xi^3 -2\xi^5) F_0 + (5\xi^2 - 3 \xi^4) G_0} \ ,
\ee
in order to avoid any division by a vanishing quantity at $\xi=0$.
The same thing is done for $J_3$.

The above expressions are valid for the nucleus and the intermediate domains,
i.e. for $l=0,\ldots, {\cal N}-2$. For the compactified domain 
${\cal D}_{\rm ext}$, the quantity to be considered is 
$r^2 \nabla f$ instead $\nabla f$. Indeed, gradients in the 
compactified domain are used in the computation of non-linear terms in
the relativistic gravitational field equations (scalar products of gradients 
of the metric potentials). We
shall see below that the source of the Poisson equation on 
${\cal D}_{\rm ext}$ is to be multiplied by $r^4$, so that if each
gradient is multiplied by $r^2$, this multiplication by $r^4$ is
automatically performed. The orthonormal components of $r^2 \nabla f$
on ${\cal D}_{\rm ext}$ are
\begin{eqnarray}
 r^2 \times \der{f}{r} & = & - \l( \der{U}{\xi} \r) ^{-1} \der{f}{\xi} \\
 r^2 \times {1\ov r} \der{f}{\theta} & = & {1\ov U} \der{f}{\theta'} -
	\l( \der{U}{\xi} \r) ^{-1}  {1\ov U}\der{U}{\theta'}
	\der{f}{\xi} \\
 r^2 \times {1\ov r\sin\theta} \der{f}{\varphi} & = & {1\ov U\sin\theta'} 
	\der{f}{\varphi'} - \l( \der{U}{\xi} \r) ^{-1}  
	{1\ov U\sin\theta'}\der{U}{\varphi'}
	\der{f}{\xi} \ . 	
\end{eqnarray}

The Laplacian of a scalar field $f$ reads
\begin{eqnarray}
 \Delta f & = & J_1^{-1} \Bigg\{ J_1^{-1} \l( 1 + J_2^2 + J_3^2 \r) 
	{\partial^2 f\ov \partial \xi^2} + {2\ov R_l} \der{f}{\xi} \Bigg\} 
	+ {1\ov R_l^2} \Delta_{\theta\varphi} f
  - J_1^{-1} \Bigg\{ 2 \Bigg( {J_2\ov R_l} 
		{\partial^2 f\ov \partial \theta\partial \xi} 
	+ {J_3\ov R_l \sin\theta} 
	{\partial^2 f \ov \partial \varphi\partial \xi} \Bigg)
				\nonumber \\
&&  + \Bigg[ {1\ov R_l^2} \Delta_{\theta\varphi} R_l + J_1^{-1}
	\Bigg( J_1^{-1} ( 1 + J_2^2 + J_3^2) 
	{\partial^2 R_l \ov\partial \xi^2}
  - 2 \Big( {J_2 \ov R_l}
		{\partial^2 R_l\ov \partial \theta\partial \xi} 
	+ {J_3\ov R_l \sin\theta} 
	{\partial^2 R_l \ov \partial \varphi\partial \xi} \Big) \Bigg)
	\Bigg] \der{f}{\xi} \Bigg\} \ ,		\label{e:laplacien}
\end{eqnarray}
where the primes on $\theta$ and $\varphi$ have been abandoned again and
the following abbreviation has been introduced:
\be	\label{e:lapang}
  \Delta_{\theta\varphi}  :=  {\partial^2 \ov \partial{\theta}^2}
	+ {1\ov \tan\theta} {\partial \ov \partial\theta}
	+ {1\ov \sin^2\theta} {\partial^2 \ov \partial{\varphi}^2} \ .
\ee

\subsection{Resolution of the Poisson equation}	\label{s:resol_Poisson}

For many astrophysical applications, one has to solve the Poisson-like
equation
\be
  \Delta f = \sigma(f) \ ,			\label{e:Poisson}
\ee
for some `potential' $f$.  Note that for relativistic computations,
$\sigma(f)$ is not compactly supported (see e.g. \cite{BonGSM93}) and
generally decreases as $1/r^4$ when $r\rightarrow +\infty$.

When expressed in terms of the variables $(\xi,\theta,\varphi)$,
the Laplacian takes the complicated form 
(\ref{e:laplacien}), for which it is not obvious to find eigenfunctions.
Therefore, we introduce, in each domain ${\cal D}_l$, a new radial 
coordinate
\be
	\zeta := \alpha_l \xi + \beta_l	\ ,
\ee
where $\alpha_l$ and $\beta_l$ are the same constants as in 
Eqs.~(\ref{e:r0=afbg}) and (\ref{e:rl=afbg}) 
(in the nucleus: $\beta_0=0$). In the exterior domain, we introduce
\be
	\eta := \alpha_{\rm ext} (\xi - 1) \ ,
\ee
where $\alpha_{\rm ext}$ is the same constant as in 
Eq.~(\ref{e:u=af}).
We may then split the Laplacian operator $\Delta$ into a
{\em pseudo-Laplacian} $\tilde\Delta$ and a part which would vanish
if the domains ${\cal D}_l$ were exactly spherical (in this case, the
coordinates $\zeta$ and $\eta$ introduced here above would coincide
with the physical coordinates $r$ and $u=1/r$ respectively).
By {\em pseudo-Laplacian}, we mean the operator which once expressed
in terms of $(\zeta,\theta,\varphi)$ has the same structure than the
Laplacian operator in spherical coordinates:
\be
 \label{e:tilde_delta}
  \tilde \Delta  :=  {\partial^2 \ov \partial\zeta^2}
	+ {2\ov \zeta} {\partial\ov \partial\zeta}
	+ {1\ov \zeta^2} \Delta_{\theta\varphi}	\ ,
\ee
where $\Delta_{\theta\varphi}$ is defined by Eq.~(\ref{e:lapang}).
In the exterior domain, the {\em pseudo-Laplacian} is defined
instead by
\be
  \tilde \Delta  :=  {\partial^2 \ov \partial\eta^2}
	+ {1\ov \eta^2} \Delta_{\theta\varphi}	\ .	
\ee
It is much easier to invert the operator $\tilde\Delta$ than
the operator $\Delta$: using spherical harmonics in $(\theta,\varphi)$,
the problem reduces to a system of second order ordinary differential
equations with respect to the variable $\zeta$. Moreover, the junction
conditions between the various domains are easily imposed, as
explained below.

The Poisson equation (\ref{e:Poisson}) becomes
\be \label{e:a_tilde_Poisson}
	a \tilde \Delta f = \sigma(f) +{\cal R}(f)	\ ,
\ee
where
\be
  a := \alpha_l^2	\, J_1^{-2}\, (1+ J_2^2 + J_3^2)	\ , 
\ee
\begin{eqnarray}
 {\cal R}(f) & := & \l[ J_1^{-1} {R_l\ov \xi + \beta_l/\alpha_l}
		 (1+ J_2^2 + J_3^2)- 1 \r] 
 	{2\ov J_1 R_l}  \der{f}{\xi}  
	+ 
	\l[ J_1^{-2} {R_l^2\ov (\xi + \beta_l/\alpha_l)^2}
		 (1+ J_2^2 + J_3^2)- 1 \r] 
	{1\ov R_l^2} \Delta_{\theta\varphi} f \nonumber \\
  &&  + J_1^{-1} \Bigg\{ 2 \Bigg( {J_2\ov R_l} 
		{\partial^2 f\ov \partial \theta\partial \xi} 
	+ {J_3\ov R_l \sin\theta} 
	{\partial^2 f \ov \partial \varphi\partial \xi} \Bigg)
				\nonumber \\
&&  + \Bigg[ {1\ov R_l^2} \Delta_{\theta\varphi} R_l + J_1^{-1}
	\Bigg( J_1^{-1} ( 1 + J_2^2 + J_3^2) 
	{\partial^2 R_l \ov\partial \xi^2}
 - 2 \Big( {J_2 \ov R_l}
		{\partial^2 R_l\ov \partial \theta\partial \xi} 
	+ {J_3\ov R_l \sin\theta} 
	{\partial^2 R_l \ov \partial \varphi\partial \xi} \Big) \Bigg)
	\Bigg] \der{f}{\xi} \Bigg\} \ .		\label{e:reste}
\end{eqnarray}
In order to let appear only the operator $\tilde \Delta$ in the
left-hand-side of Eq.~(\ref{e:a_tilde_Poisson}), we introduce
\be
	a_l := \max_{{\cal D}_l} a	 \ ,	
\ee
and recast Eq.~(\ref{e:a_tilde_Poisson}) into
\be \label{e:tilde_Poisson}
	\tilde\Delta f = {1\ov a_l} \l[
	\sigma(f) + {\cal R}(f) + (a_l - a) \tilde\Delta f \r] \ .
\ee
Since $f$ appears on the right-hand-side of this equation, we solve it
by iteration.
Besides, we introduce some relaxation in the computation of the term
$\tilde\Delta f$ in the right-hand-side of Eq.~(\ref{e:tilde_Poisson}).
More specifically, we solve at each step of the iterative scheme
the following equation
\be \label{e:tilde_Poisson_res}
	\tilde\Delta f^{J+1} = \tilde \sigma^J	\ ,
\ee
where the index $J$ denotes the step at which the quantities are taken
and $\tilde\sigma^J$ is the following source, computed from the
value of $f$ at the step $J$:
\begin{eqnarray}
  \tilde \sigma^J & = & a_l^{-1} \Big\{ \sigma(f^J) + {\cal R}(f^J)
 + (a_l - a) \l[ \lambda \tilde\sigma^{J-1} + (1-\lambda)
	\tilde\sigma^{J-2}	\r] \Big\}	\ .
\end{eqnarray}
In this expression, $\lambda$ is a relaxation parameter (a typical value
is $\lambda = 1/2$) and ${\cal R}(f^J)$ is to be computed according to 
Eq.~(\ref{e:reste}).
For the first step ($J=0$), $f^J$, $\tilde\sigma^{J-1}$
and $\tilde\sigma^{J-2}$ are set to zero or to their value at a previous 
step in an evolutionary scheme.

We have exposed the method of resolution of Eq.~(\ref{e:tilde_Poisson_res})
elsewhere \cite{BonM90,BonGSM93}.
Let us simply mention that we first perform 
a transformation from the bases in $(\theta,\varphi)$
described in  Sect.~\ref{s:expansion} (Chebyshev polynomials in $\cos\theta$ 
-- Fourier expansion in $\varphi$) to spherical harmonics
$Y_\ell^m(\theta,\varphi)$, by means of a matrix multiplication onto the
coefficients of the $\theta$ expansion. For each value of $(\ell,m)$,
Eq.~(\ref{e:tilde_Poisson_res}) gives then a second order ordinary
differential equation with respect to $\zeta$, the solution of which
amounts to invert a banded matrix. Two solutions of the homogeneous
equation ($\Delta f = 0$) are then added in order to connect 
the solution and its
first derivative across the boundaries between the $\cal N$ domains.
More precisely, the global boundary condition, 
generally $f\rightarrow 0$ when
$r\rightarrow +\infty$, is imposed by setting the value of $f$ at the
exterior boundary of the external domain, which is exactly
$r=+\infty$ as explained in Sect.~\ref{s:map_externe}. 
The matching between the various domains amounts then to the resolution
of a simple system of $2{\cal N}-1$ linear equations for the 
coefficients of the homogeneous solutions to be added in each domain.
Note that this matching is
performed for each value of $(\ell,m)$.

\section{Regularization of the source of Poisson equation}
\label{s:regul} 

\subsection{Description of the method}

The analytical properties of the source of the gravitational field at
the boundary of the star depend on the equation of state (EOS). For a
polytrope of adiabatic index $\gamma$ ($P\propto n^\gamma$), the matter
density $n$ behaves as $H^{1/(\gamma-1)}$ where $H$ is the specific enthalpy.
Consequently, for $\gamma > 2$, the derivative $dn/dH$ has an infinite
value for $H=0$, i.e. at the surface of the star and $dn/dr$ diverges
at surface of the star. For values of $\gamma < 2$ only derivatives of
higher order diverge (actually there exists some value of $\gamma$,
e.g. $\gamma = 4/3$, for which all derivatives vanish or have a
finite value at the surface of the star).

In a steady state configuration $H$ is Taylor expandable at the
neighbourhood of the star's surface (this can be easily seen on
Eq.~(\ref{e:integ_prem}) below). Therefore $H$ vanishes as $r -
R(\theta,\varphi)$ where $R(\theta,\varphi)$ is the equation of the star's
surface and $n$ behaves as $n \sim [r-R(\theta, \varphi)]^{1/(\gamma
-1)}$ (this analysis remains valid even for EOS more general than the
polytropic one if $\gamma $ is defined as $\gamma =
d(\ln(P)/d\ln(H)|_{H=0}$). Consequently $n$ is generaly not a ${\cal
C}^\infty$ function. This singular behaviour implies that the ${\cal
L}^2$ truncation error of the spectral approximation is no more
evanescent and moreover that Gibbs phenomenon is present. This fact is
especialy awkward when studying the stability of equilibrium
configurations or looking for bifurcation points because high accuracy
is required. In practice $\gamma$ cannot be larger than $3$
\cite{BonFG96,BonFG98}. Note that in the litterature the potential in
spherical coordinates is often computed by expanding the source in
spherical harmonics ${Y_l}^m(\theta,\varphi)$ and by computing the
radial part with a finite difference method. In this case the Gibbs
phenomenon will appear in the angular part of the solution. The
situation is even worse if the radial part of the potential is computed
with a spectral method. A method to recover spectral accuracy in such
cases is as follows. 

We first introduce a known potential $\Phi_{\rm div}$ such that $n_{\rm div} :=
\Delta \Phi_{\rm div}$ as the same pathological behaviour as $n$ and such
that $n-n_{\rm div}$ is a regular function (at least more regular then $n$)
and numericaly solve
\be
\Delta \Phi_{\rm regu} = n - n_{\rm div}
\ee
where $\Phi_{\rm regu} := \Phi - \Phi_{\rm div}$.

Consider for instance 
\be
 \label{e:phi_div}
\Phi_{\rm div}= F(\xi,\theta,\varphi) (1-\xi^{2})^{(\alpha+2)}
\ee
where $\alpha=1/(\gamma-1)$, $F$ is an arbitrary regular function and
$\xi$ is a new radial variable such that
$\xi=1$ at the surface of the star (see Sect.~\ref{s:mapping}). It is
easy to see that $\Delta \Phi_{\rm div}$ has a term vanishing at the
surface as $(1-\xi)^{\alpha}$ (i.e. with the same pathological
behaviour as $n$). We have indeed
\begin{eqnarray}
\tilde{\Delta}\Phi_{\rm div} & = & \tilde{\Delta} F
\xi^{2}(1-\xi^{2})^{(\alpha+2)}
-4(\alpha+2)\xi(1-\xi^{2})^{(\alpha+1)}\partial_{\xi}F \nonumber \\ &
& +(\alpha+2) \left[-6(1-\xi^{2})^{(\alpha+1)}+
4(\alpha+1)\xi^{2}(1-\xi^{2})^{\alpha} \right] F \label{e:lap} 
\end{eqnarray}
where $\tilde{\Delta}$ is the Laplacian computed with respect
to ($\xi,\theta,\varphi$) (cf. Eq. (\ref{e:tilde_delta})).

The choice of the factor of $(1-\xi)^{(\alpha+2)}$ is done in order
that $\Phi_{\rm div}$ has the required regularity properties at $\xi=0$ and
the required behaviour at the boundary of the star.  The choice of
$F(\xi,\theta,\varphi)$ is arbitrary. If we choose for
$F(\xi,\theta,\varphi)$ an harmonic function, $\tilde{\Delta}F = 0$, the
first term of the r.h.s. of the Eq.~(\ref{e:lap}) vanishes. This is an
advantage because this term can be quite large and, consequently, give
rise to a large error in computing $\Phi_{\rm regu}$. We write
$\Phi_{\rm div} = \sum_{l,m}a_{lm}\Phi_{lm}$, where
\be
\Phi_{lm} := \xi^{l} (1-\xi^{2})^{(\alpha+2)} {Y_{l}}^m (\theta,\varphi) 
\ee
and where $a_{lm}$ are some numerical coefficients to be determined.
We then obtain $ n_{\rm div}(\xi,\theta,\varphi) = \sum_{l,m}a_{lm}
C_{l}(\xi) {Y_l}^m (\theta,\varphi) $, with
\be 
C_{l}(\xi)
	:= (\alpha+2)\left[-(4l+6)(1-\xi^2)^{(\alpha+1)}\xi^{l} +4(\alpha+1)\xi
		^{(l+2)}(1-\xi^{2})^{\alpha}\right] \ .
\ee 
We have now to determine the values of the coefficients $a_{lm}$ which
give the most regular function $ n_{\rm regu} := n - n_{\rm div}$. The
criterion which seems to give the best results is the following one. 

We expand $n$ and $n_{\rm div}$ as truncated series of spherical harmonics
${Y_l}^m(\theta,\varphi)$ and Chebyshev polynomial $T_{i}(\xi)$:
\be 
 n(\xi,\theta,\varphi) = \sum_{i,l,m=0}^{I,L,M} n_{ilm} \,
	T_{i}(\xi) \, {Y_l}^m (\theta,\varphi) 
\ee
and each of the functions $C_{l}(\xi)$ in a Chebyshev series:
\be 
 C_{l}(\xi) = \sum_{i}^{I} C_{li} \, T_{i}(\xi) \ .
\ee
The value of $a_{lm}$ is computed in such a way  that the $I^{th}$
coefficient of the truncated series of $n_{\rm regu}$ vanishes:  
\be
a_{lm}=n_{Ilm}/C_{lI} \ .
\ee

By means of the above procedure, we eliminate in $n$ the pathological
term vanishing as $(1-\xi)^{\alpha}$ but we introduce another
pathological term $\propto (1-\xi^{2})^{\alpha+1}$.  However the
divergence occurs in a higher order derivative of this term so that it
has a much weaker effect on the accuracy of the result. The method can
be improved by taking
\be
 \label{e:def_k}
\Phi_{\rm div} = F(\xi,\theta,\varphi)
(1-\xi^{2})^{\alpha+2}
	\left[a_1 + a_2 (1 - \xi^2) + a_3 (1-\xi^2)^2 + \cdots
	+ a_K (1-\xi^2)^{K-1} \right] \ .  
\ee 
instead of (\ref{e:phi_div}).
The coefficients $a_{k}$ are chosen in a such a way that the $1^{st}$,
$2^{th}, \ldots, K^{th}$ derivatives of $n_{\rm regu}$ vanish at $\xi=1$. 
Let us call $K$ the {\em regularization degree} of the procedure. 

Note that, since $\Phi_{\rm div}$ and $\partial_{\xi} \Phi_{\rm div}$ vanish at
the surface of the star, the boundary condition one has to impose to
solve $\Delta \Phi_{\rm regu} = n_{\rm regu}$ is the same than that for $\Delta
\Phi = n$. We want to point out that, the above regularization
technique can be used, {\it mutatis mutandis}, also when a finite
difference method is used.

\subsection{Examples}

\begin{figure}
\centerline{\epsfig{figure=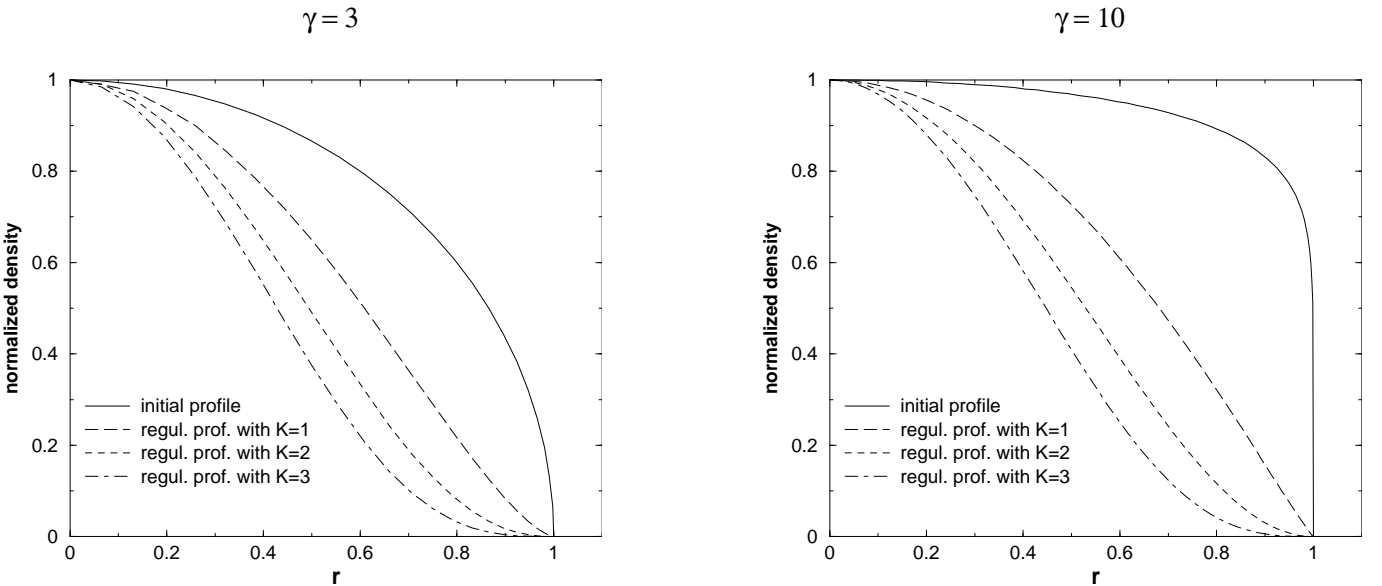,height=8cm}}
\caption[]{\label{f:gam30} Original and regularized density profiles for
$\gamma=3$ and $\gamma=10$ polytropes. 
The regularized profiled are rescaled to take the value $1$ at the origin.}
\end{figure}

Consider two polytropic EOS of adiabatic index $\gamma=3$ and
$\gamma=10$ with a spherically symmetric distribution of the enthalpy
$H=1-\xi^2$. The corresponding sources density are $n_3(\xi) =
(1-\xi^2)^{1/2}$ and $n_{10}(\xi) = (1-\xi^{2})^{1/9}$.
Figure~\ref{f:gam30} shows the mass distributions $n$ and $n_{\rm regu}$
for various values of the regularization degree $K$ [Eq.~(\ref{e:def_k})]. 
Note that in the case of $\gamma = 10$ 
the procedure improves considerably the behaviour of the source 
$n_{\rm regu}$ even for $K=1$.

\begin{figure}
\centerline{\epsfig{figure=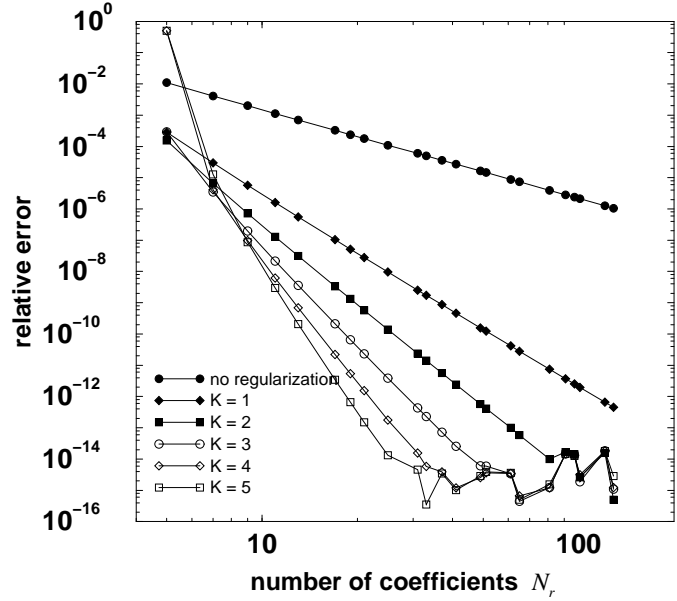,height=8cm}}
\caption[]{\label{f:erreur_regu} Relative ${\cal L}^1$
error $\epsilon$ on the gravitational
field as a
function of the number of degrees of freedom $N_r$ for different regularization
degrees $K$.}
\end{figure}

\begin{figure}
\centerline{\epsfig{figure=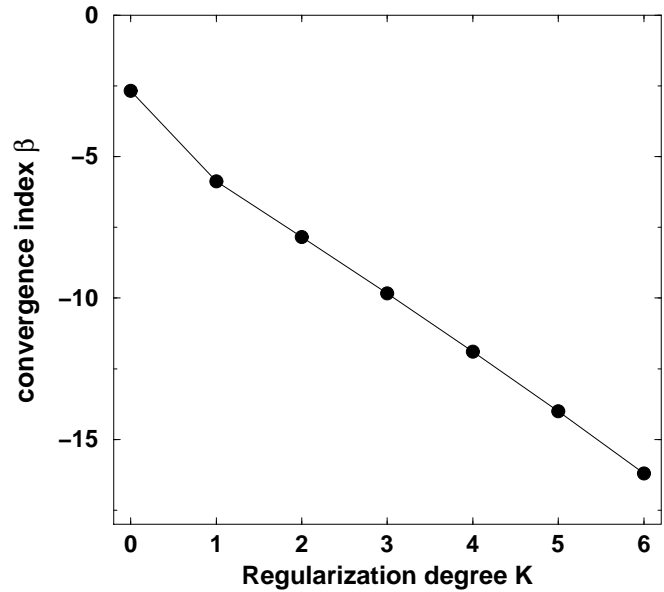,height=8cm}}
\caption[]{\label{f:pente} Dependence of the exponent $\beta$ on the
regularization degree $K$.} 
\end{figure}

The method can be tested in the case of $\gamma=3$ by direct comparison
with the analytical solution. In this case the gravitational field
$G = \partial_{r} \Phi$ reads
\be  
G = \frac{1}{r^{2}}\int_{0}^{r}(1-u^{2})^{1/2}u^{2} du
  =
\frac{1}{8r^{2}}\lbrack\arcsin{r} + r (1-r^{2})^{1/2}-2r(1-r^{2})^{3/2}
\rbrack 
\ee 
Figure~\ref{f:erreur_regu} shows the relative ${\cal L}^1$
error $\epsilon$ on $G$ as a
function of the number of degrees of freedom $N_r$. The error $\epsilon$
follows approximately a power law $\epsilon \propto N_r^{-\beta}$.
The dependence of the exponent $\beta$ with respect to the regularization
degree $K$ is shown in Figure~\ref{f:pente}. 
A value as high as of $\beta \approx
17$ can be achieved with only $K = 6$.  Note that the relation
$\epsilon = N_{r}^{-\beta}$ is only an approximate law. This means that
the error tends to become evanescent when the regularization degree
increases.

\section{ILLUSTRATIVE APPLICATIONS} \label{s:illustr}

\subsection{3-D stationary configurations} \label{s:code}

In this section, we sketch the general structure of a code for computing
single star stationary configurations under the influence of rotation and 
a tidal potential. For simplicity we present
only the Newtonian case, the relativistic one showing no new qualitative
feature but simply involving more equations.

The equilibrium configuration of a cold star rotating rigidly at 
the angular velocity $\Omega$ with respect to some inertial frame and
embedded in a tidal potential $\Phi_{\rm tide}$ is governed by the following 
three 
equations\footnote{see \cite{BonG97} for a discussion of 
these equations, including the relativistic case.}
\be \label{e:Poisson_grav}
	\Delta \Phi_{\rm grav} = 4\pi G\, \rho \ ,
\ee
\be \label{e:integ_prem}
	H + \Phi_{\rm grav} -{1\ov 2} (\Omega r \sin\theta)^2
		+ \Phi_{\rm tide} = {\rm const}	\ , 
\ee 
\be \label{e:eos}
	\rho = \rho(H)	\ .
\ee
Equation~(\ref{e:Poisson_grav}) is the Poisson equation linking 
the gravitational potential $\Phi_{\rm grav}$ to 
the mass density $\rho$.
Equation~(\ref{e:integ_prem}) is the first integral which can be derived
from the Euler equation governing the (perfect) fluid velocity
under the stationarity assumption; this equation relates the specific
enthalpy of the fluid $H$ to the internal and external potentials. 
Finally Eq.~(\ref{e:eos}) is the 
matter equation of state in the zero-temperature approximation.

The number of domains used for solving this problem
is ${\cal N} = 3$ : one domain ${\cal D}_0$ for the star
(the nucleus, cf. Sect.~\ref{s:map_nucl}), one intermediate domain 
${\cal D}_1$ (cf. Sect.~\ref{s:map_interm}), 
the outer boundary of which is spherical and the external
domain ${\cal D}_2$ (cf. Sect.~\ref{s:map_externe}).
In fact, if one would have to treat only the Newtonian case, one domain
would be sufficient (i.e. the nucleus) because Eq.~(\ref{e:Poisson_grav})
has a compact support, which is no more true in the relativistic case.

A solution is specified by the central value of $H$ (or $\rho$),
$H_{\rm c}$ say,
the value of $\Omega$ and the expression of $\Phi_{\rm tide}$.
These quantities being given, the iterative method of resolution is as 
follows. 
The $\cal N$ domains are first taken to be exactly spherical.  
One starts from a very crude density profile, for instance 
$\rho = {\rm const}$ in ${\cal D}_0$. Solving the Poisson 
equation~(\ref{e:Poisson_grav}) by means of the method presented in 
Sect.~\ref{s:resol_Poisson} leads to the gravitational potential
$\Phi_{\rm grav}$. Inserting its value into Eq.~(\ref{e:integ_prem}) give
a new profile for the specific enthalpy $H$ (the constant on the
right-hand-side of Eq.~(\ref{e:integ_prem}) is fully determined by 
the requirement that $H=H_{\rm c}$ at the centre of the star).
The surface of the star being defined by $H=0$, its equation
$r=S_0(\theta,\varphi)$ [using the notation of Eq.~(\ref{e:S_l:def})]
is found by searching for the equipotential $H=0$ in the newly determined
$H$ field. This defines a new domain ${\cal D}_0$. The corresponding
mapping $R_0(\xi,\theta,\varphi)$, i.e. the value of the constant
$\alpha_0$ and the functions $F_0(\theta,\varphi)$ and
$G_0(\theta,\varphi)$ [cf. Eq.~(\ref{e:r0=afbg})] is computed according to 
the procedure
described in Sect.~\ref{s:map_nucl}. The new intermediate domain 
${\cal D}_1$ is defined by the new inner boundary ${\cal S}_0$ (the
surface of the star) and the unchanged spherical outer boundary
${\cal S}_1$. The corresponding mapping 
$R_1(\xi,\theta,\varphi)$ is computed according to the procedure described 
in Sect.~\ref{s:map_interm}. The external domain ${\cal D}_2$
remains unchanged. 

The physical location $(r,\theta,\varphi)$ of the collocation
points $(l,\xi_i,\theta_j,\varphi_k)$ (where $l$ is the domain index)
corresponding to these new mappings is a priori different from that of
the previous mappings, where all the fields were known. Therefore,
one has to compute the values of the fields at the new collocation points.
In the present case, it is sufficient to do so only for the
specific enthalpy $H$.
In the domain no. $l$, the collocation point $(\xi_i,\theta_j,\varphi_k)$
has the physical radial coordinate
\be
	r = R_l^J(\xi_i,\theta_j,\varphi_k)	 \ ,
\ee
where the superscript $J$ refers to the step in the iterative procedure:
$R_l^J(\xi,\theta,\varphi)$ is the current value of the mapping of
the domain ${\cal D}_l$, whereas $R_l^{J-1}(\xi,\theta,\varphi)$
is the previous value. Let us denote by 
$(L^{J-1}(r,\theta,\varphi),\Xi^{J-1}(r,\theta,\varphi))$ the inverse
mapping at the previous step. This inverse mapping is computed by 
searching for the zero of the function 
$(l,\xi)\mapsto r - R_l(\xi,\theta,\varphi)$. 
The values of $H$
at the collocation points of the new mapping are given by
\begin{eqnarray}
  H^J(l,\xi_i,\theta_j,\varphi_k) = H^{J-1}\Big( &&
	L^{J-1}(r^J,\theta_j,\varphi_k),
  \Xi^{J-1}(r^J,\theta_j,\varphi_k),\theta_j,\varphi_k \Big)	
		\label{e:H_new} \ ,
\end{eqnarray}	
where $r^J := R_l^J(\xi_i,\theta_j,\varphi_k)$.
The value of $H$ on the right-hand-side is to be taken at a point which
a priori does not coincide with a collocation point in $\xi$.
It is computed by a direct summation, by means the spectral expansion
of $H$. Using the notations of Sect.~\ref{s:expansion}, it writes
\begin{eqnarray}
  H(l,\xi,\theta,\varphi) & = &  
	\sum_{k=0}^{N_\varphi-1} \Bigg[
	\sum_{j=0}^{N_\theta-1} \l(
	\sum_{i=0}^{N_r-1} \hat H_{lkji} X_{kji}(\xi) \r) \, 
  \Theta_{kj}(\theta)	\Bigg]
	\Phi_k(\varphi)	\ ,	\label{e:interpol}
\end{eqnarray}
where the $\hat H_{lkji}$ are the coefficients of $H$ is the domain
no. $l$. Note that from the computational point of view, this summation
is the most expensive operation of the method: it scales indeed
as $(N_r N_\theta N_\varphi)^2$. It may be possible to replace
the whole summation (\ref{e:interpol}) by a truncated one
or by some interpolation from the values of $H$ at the collocation points, in
order to reduce the computational cost. 
The main advantage of the summation (\ref{e:interpol}) is that it does
not introduced any additional error in the method: the right-hand-side
of Eq.~(\ref{e:interpol}) is the value of $H$ at the specified point
within the spectral accuracy.

Once $H$ is computed at the collocation points of the new mapping
by means of Eq.~(\ref{e:H_new}), the equation of state (\ref{e:eos})
is used to find the values of the mass density $\rho$ at the
collocation points. A new iteration may then begin.

In all the computations we have made, we have found that this procedure
converges. For stationary rotating stars in general relativity,
a rigorous proof of convergence of such iterative method (except for
the re-mapping of the physical space at each step) has been given by
Schaudt \& Pfister \cite{SchP96}.

\subsection{MacLaurin ellipsoids} \label{s:MacLaurin}

The multi-domain spectral method can handle constant density
(incompressible matter) rotating bodies without any Gibbs phenomenon.
With classical spectral methods, the Gibbs phenomenon would have been
very severe since the density itself, and not some of its derivatives,
is discontinuous across the stellar surface for incompressible
fluids.  This gives us the opportunity to quantify the accuracy of the
method since exact analytical solutions are known for incompressible
bodies:  the so-called {\em ellipsoidal figures of equilibrium} (see
e.g. \cite{Cha69}). Note that an ellipsoid is not a particular case for
the mapping (\ref{e:r0=afbg}): all the coefficients of the expansion of
$F_0(\theta,\varphi)$ and $G_0(\theta,\varphi)$ onto the bases
described in Sect.~\ref{s:expansion} are non-zero.  In this respect,
the ellipsoidal figures constitute a strong test of the method.

\begin{figure}
\centerline{ \epsfig{figure=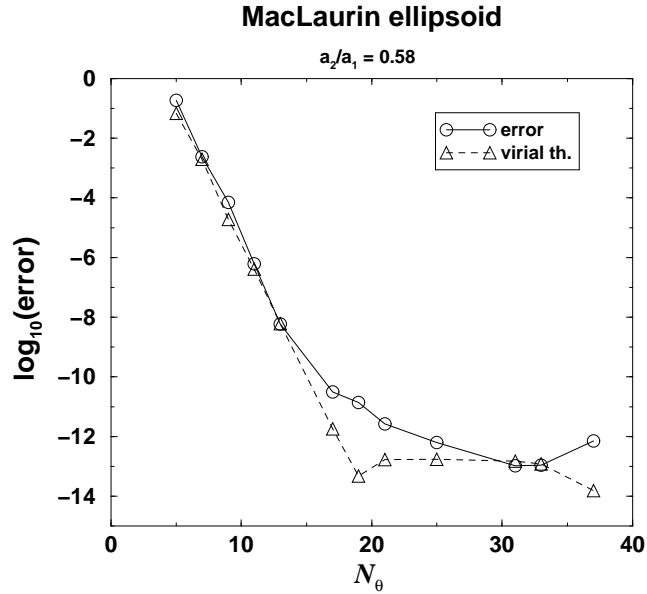,height=8cm} }
\caption[]{\label{f:virielml} 
Logarithm of the relative error of the numerical solution
with respect to the number of degrees of freedom in $\theta$ 
for a MacLaurin spheroid at the 
Jacobi-Dedekind bifurcation point (the number of degrees of freedom in $r$ is
$N_r = 2N_\theta -1$). 
Also shown is the error in the
verification of the virial theorem.}  
\end{figure}

For single rigidly rotating objects in the 
Newtonian regime, the more simple ellipsoidal solutions are constituted by 
the family of {\em MacLaurin spheroids}, which are axisymmetric about their
rotation axis. We have computed them by means of the procedure
presented in Sect.~\ref{s:code}, setting $\Phi_{\rm tide} = 0$
and the equation of state (\ref{e:eos}) to be simply
$\rho = {\rm const}$. The axisymmetry allows to employ
$N_\varphi = 1$.
The code converges towards ellipsoidal configurations
and we measure the error by comparing the eccentricity 
$e := \sqrt{1 - (r_{\rm p}/r_{\rm eq})^2}$ (where $r_p$ and
$r_{\rm eq}$ are respectively the polar and equatorial radii) of the
numerical solution with that of the analytical solution. 
The result of this comparison is presented in Fig.~\ref{f:virielml}
for a MacLaurin spheroid located on the MacLaurin sequence at the
point where the Jacobi and Dedekind sequences branch off: the eccentricity
is $e = 0.8127$, which corresponds to the ratio 
$r_{\rm p}/r_{\rm eq} = 0.5827$. 
Shown in Fig.~\ref{f:virielml} is the relative error on the eccentricity
as a function of the number of coefficients in the $\theta$ expansion.
For these calculations, the number of coefficients in the $\xi$ 
expansion in each domain is $N_r = 2 N_\theta - 1$. 
The straight line behaviour of the 
left side of Fig.~\ref{f:virielml} 
shows that the error is {\em evanescent}, i.e. that it decreases as 
$\exp(-N_\theta)$. For $N_\theta \gta 20$, the error saturates at the
level of $10^{-12} - 10^{-11}$. This is due to the round-off errors in
the computation, which is performed with a 15-digit accuracy.
It is instructive to compare this result with that obtained with 
a classical spectral method, i.e. with fixed spherical
domains, as exposed in \cite{BonGSM93}. For instance, 
the Fig.~\ref{f:virielml} can be directly compared with Fig.~7 of 
\cite{BonGSM93}: this latter shows an power-law error decay only
(of the type $N_\theta^{-4.5}$), due to the Gibbs phenomenon at
the star's surface. Moreover, the error saturates at the level of
$10^{-5}$. Note that this result was obtained with a polytropic equation
of state (adiabatic index $\gamma = 2$), for which the density is continuous
across the surface of star; the fixed-spherical-domain spectral method 
presented in \cite{BonGSM93} was not able to treat incompressible
fluid.

Also shown in Fig.~\ref{f:virielml} is the relative accuracy with which 
the 3-D virial theorem is satisfied. The 3-D
Newtonian virial theorem\footnote{as opposed
to the 2-D virial identity, see \cite{BonG94a} and \cite{BonG94b} 
for a discussion} states that for a stationary configuration
$2T + 3 P + W = 0$, where $T$ is the total kinetic energy (with respect
to the inertial frame), $P$ is the integral of the pressure throughout the
star and $W$ is the gravitational potential energy. We have computed each
of these three integrals for the numerical solution and evaluated the
quantity
\be
	\varepsilon := \l| 1 - {2T + 3P \ov |W|} \r| \ .
\ee
For an exact solution, $\varepsilon = 0$. The triangles plotted in
Fig.~\ref{f:virielml} depict the value of $\log_{10}\varepsilon$ for the
numerical models. Figure~\ref{f:virielml} shows that the virial error
is very well correlated with the error evaluated by a direct comparison
with the analytical solution. This gives us a strong confidence when
using the virial error to evaluate the numerical error in more
general cases, when no analytical solution is available.

\subsection{Roche ellipsoids}

Roche ellipsoids are equilibrium solutions for incompressible 
fluid bodies in a synchronized binary system, within the approximation of
taking only the second order term in the expansion (around the center of
mass of one star) of the gravitational potential of the companion.
They are obtained by setting
\be
  \Phi_{\rm tide} = - { G M_{\rm comp} \ov |a|} \l( 1 + {x\ov a} 
	+ {2x^2 - y^2 -z^2\ov a^2} \r)
\ee
in Eq.~(\ref{e:integ_prem}), where $a$ is the abscissa of the center of mass
of the companion in the Cartesian frame $(x,y,z)$ centered at the
center of mass of the star under consideration. 
Note that in Eq.~(\ref{e:integ_prem}),
$r$ must now be the distance to the center of mass of the binary system
and $\theta$ the angle with respect to the rotation axis of the system. 
Moreover, $\Omega$ must be chosen so that $\Omega^2 = G(M + M_{\rm comp})/a^3$,
in order that the linear term in $x$ which appear in Eq.~(\ref{e:integ_prem})
vanishes and one is left with an ellipsoidal solution. 

The analytical solutions for Roche ellipsoid are given in the classical
book by Chandrasekhar \cite{Cha69}. However they are given with an
accuracy of five digit only (Table~XVI in Ref.~\cite{Cha69}), which 
is not sufficient for our comparison project: the accuracy achieved by
the numerical code is far better than $10^{-5}$ as we shall see below. 
Therefore, we have written a small Mathematica \cite{Wol92}
program to compute
Chandrasekhar's ``index symbol'' $A_1$, $A_2$ and $A_3$ and obtain
Roche solutions with an arbitrary number of digits. 

\begin{figure}
\centerline{ \epsfig{figure=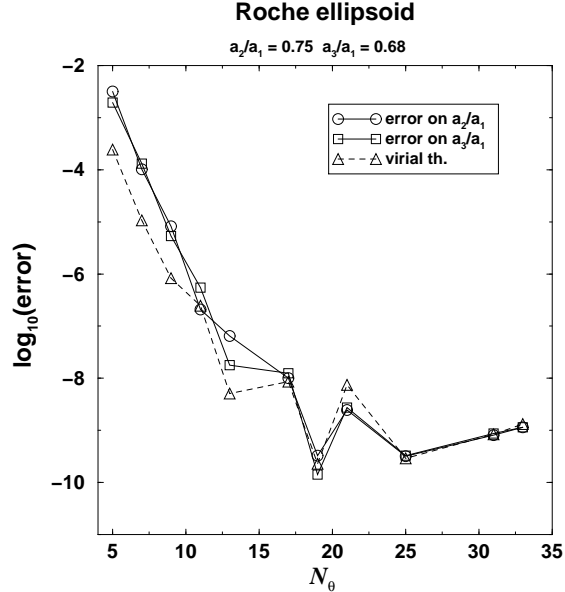,height=8cm} }
\caption[]{\label{f:virielro} 
Logarithm of the relative global error of the numerical solution
with respect to the number of degrees of freedom in
$\theta$ for a Roche ellipsoid for an equal mass binary
system and
$\Omega^2/(\pi G\rho) = 0.1147$ (the numbers of degrees of freedom in the
other directions are $N_r = 2N_\theta-1$ and $N_\varphi = N_\theta -1$). 
 Also shown is the error in the
verification of the virial theorem.}  
\end{figure}

\begin{figure}
\centerline{ \epsfig{figure=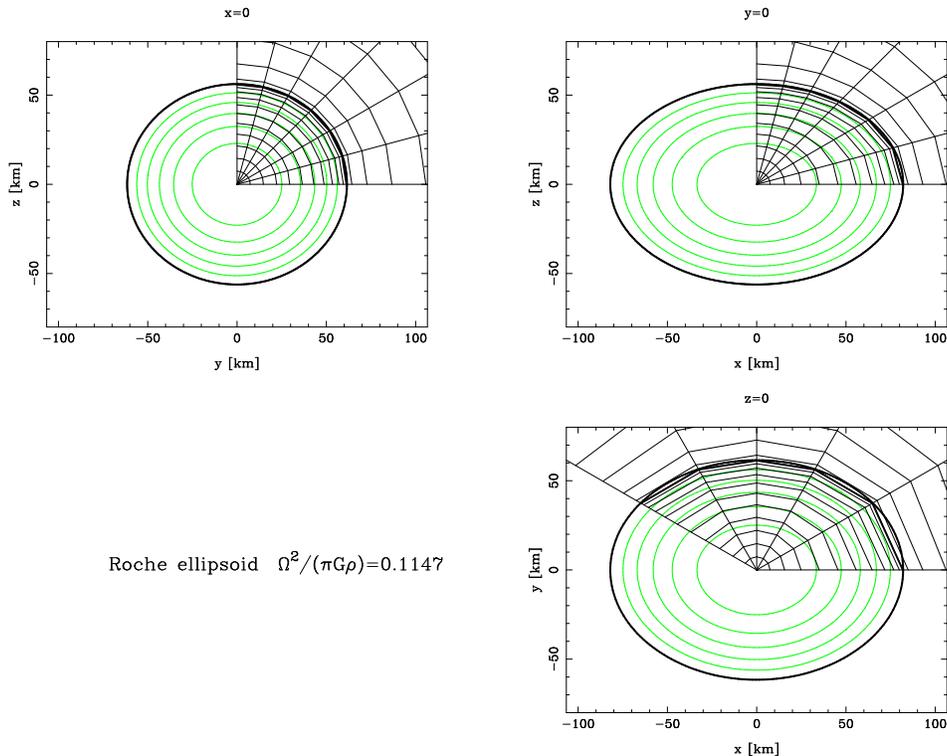,height=10cm} }
\caption[]{\label{f:coupe_roche} 
Orthogonal plane sections in 
the numerical solution obtained for the Roche ellipsoid represented by
the second set of points starting from the left on Fig.~\ref{f:virielro}  
(i.e. corresponding to $N_r = 13$, $N_\theta = 7$ and $N_\varphi = 6$). 
Shown are the iso-enthalpy lines, as well as the numerical grid.
This computation took a few seconds on a R4400/150MHz processor.
} 
\end{figure}

Figure~\ref{f:virielro} present the results of the comparison between
the numerical solution obtained by means of the method described in 
Sect.~\ref{s:code} and the analytical solution. Let us recall 
that ellipsoidal shapes are not privileged in our formalism, so that
this type of comparison constitute a strong test of our method. 
The comparison is conducted at fixed values of $\Omega/\rho$ and the
mass ratio $M_{\rm comp}/M$. 
Two global errors can be then defined: (i) the error on the axis ratio
$a_2/a_1$ and (ii) the error on the axis ratio $a_3/a_1$, $a_1$ being 
the major axis of the triaxial ellipsoid (directed along the line of the
two centers of mass), $a_2$ being the orthogonal axis in the orbital plane,
and $a_3$ being the axis perpendicular to the orbital plane. 
These two errors are shown in Figure~\ref{f:virielro} for a Roche ellipsoid
with $\Omega^2/(\pi G\rho) = 0.1147$ and $M_{\rm comp}/M = 1$. The 
corresponding axis ratios are $a_2/a_1 = 0.7506$ and $a_3/a_1 = 0.6853$. 
The numerical solution is depicted in Fig.~\ref{f:coupe_roche} by 
three plane sections obtained with the following (small) 
numbers of coefficients: 
$N_r = 13$, $N_\theta = 7$ and $N_\varphi = 6$. Also show in this Figure
is the numerical grid (collocation points) used in the problem (only the
domain ${\cal D}_0$ and a part of ${\cal D}_1$ are represented in the 
Figure). Even with such a small number of points, the relative error
is of order $1\times 10^{-4}$ (cf. Fig.~\ref{f:virielro}) ! This explains
why despite the numerical grid is quite coarse, the iso-enthalpy
surfaces shown in Fig.~\ref{f:coupe_roche} are so smooth.  

Figure~\ref{f:virielro} gives the two global errors as a functions of the
number of coefficients in the $\theta$ expansions, $N_\theta$. The
number of coefficients employed in the other directions are 
$N_r = 2 N_\theta - 1$ and $N_\varphi = N_\theta - 1$. As in 
Fig.~\ref{f:virielml}, the exponential decay of the error for
$N_\theta \lta 13$ means that the error is evanescent. For \
$N_\theta\gta 19$, the error saturates at the level of a few $10^{-10}$
due to the round-off errors in the computation, this latter being performed
with a 15-digit accuracy. The cost in CPU time for different numbers of
degrees of freedom is shown in Table~\ref{t:prix_a_payer}.

\begin{table}
\caption[]
{ \label{t:prix_a_payer} 
CPU time cost on a R4400/150MHz processor as a
function of the number of degrees of freedom for the calculation of the Roche 
ellipsoid configuration corresponding to Fig.~\ref{f:virielro}. 
The iteration is halted when the relative discrepancy between two 
successive steps reaches $10^{-13}$.
}
\begin{tabular}{ccccc}
$N_r$	& $N_\theta$ & $N_\varphi$	& \# of steps	& CPU time per step (s) \\
\tableline
25	& 13	& 12	& 116	& 6.92 \\
33	& 17	& 16	& 107	& 24.2\\
49	& 25	& 24	& 115	& 189.16 \\
65	& 33	& 32	& 106	& 861.6 \\
\end{tabular}
\end{table}

Also shown in Fig.~\ref{f:virielro} is the relative accuracy with which the
3-D virial theorem is satisfied. This error estimator is defined in the same
way as in Sect.~\ref{s:MacLaurin}. As in the axisymmetric case (MacLaurin
ellipsoids), we find a high correlation between the virial error and the
errors obtained by direct comparison with the analytical solution.

\section{Conclusion and perspectives} \label{s:conclu}

We have presented a new numerical approach capable of handling the
surface discontinuities of stellar configurations, provided these
discontinuities are star-like, which covers a wide range of astrophysically
relevant situations. When used along with spectral methods
this adaptive-domain technique ensures that no Gibbs phenomenon can appear.
This results in a very high precision (evanescent error), as demonstrated
in Sect.~\ref{s:illustr} by comparison with exact analytical solutions. 
The relative error for 3-D configurations can reach $10^{-10}$ with a
relatively small number of degrees of freedom 
($N_r\times N_\theta \times N_\varphi = 37\times 19\times 18$ in each domain).
Let us recall that very high accuracy is required for a lot of
astrophysical problems such as numerical stability analysis. Among these
problems let us mention the study of symmetry breaking of rapidly
rotating stars and the determination of the orbital frequency of the
last stable orbit of a neutron stars binary system.

\begin{figure}
\centerline{\epsfig{figure=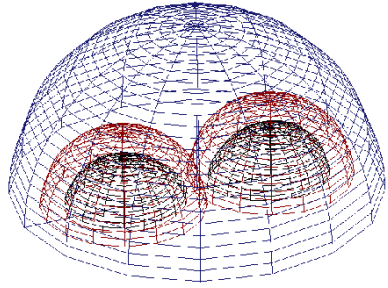,height=4cm}}
\caption[]{\label{f:m-grille}
Representation of the numerical domains that we use to compute
relativistic steady-state configurations of binary neutron stars systems. The
external domain extends to spatial infinity in order to compute the exact
gravitational potentials. Due to the symmetry of the problem, only the $z > 0$
part of space is taken into account.
}
\end{figure}

The multi-domain spectral method is particularly well adapted to the
computation of relativistic binary neutron star system. Three sets of
domains can be used in this problem (see Fig.~\ref{f:m-grille}): two
sets of (three or more) domains centered on each star and a third set
of (two or more) domains centered at the intersection between the
rotation axis and the orbital plane. This latter set of domains which
reaches spatial infinity is required to compute the gravitational field
of relativistic configurations. When needed, the quantities computed on
one of the three domain sets are evaluated at the collocation points of
another set by means of the method presented in Sect.~\ref{s:code}.  We
are currently applying this numerical method to the computation of
steady-state configurations of relativistic counter-rotating (i.e.
irrotational with respect to an inertial frame) neutron star binaries,
following the formulation developed in \cite{BonGM97a}. We will report
on the astrophysical results in a forthcoming paper.

An interesting  byproduct of the present technical paper is the
following one.  In a previous work \cite{BonGSM93}, we had been able to
demonstrate that the virial error is representative of the true error
(measured by direct comparison with analytical solutions) only in the
spherically symmetric case. We had inferred that this remains valid in
the axisymmetric and 3-D cases. In the present work, we have confirmed
this conjecture, thanks to the ability of the present method to treat
incompressible fluids, for which 3-D analytical solutions are
available.

\end{document}